\documentstyle[epsf]{cupconf}
\ifoldfss
\else
  \ifnfssone
    \newmathalphabet{\mathit}
      \addtoversion{normal}{\mathit}{cmr}{m}{it}
      \addtoversion{bold}{\mathit}{cmr}{bx}{it}
    \newmathalphabet{\mathcal}
      \addtoversion{normal}{\mathcal}{cmsy}{m}{n}
    \else
    \ifnfsstwo
    \fi
  \fi
\fi

\def\hexnumber#1{\ifcase#1 0\or1\or2\or3\or4\or5\or6\or7\or8\or9\or
 A\or B\or C\or D\or E\or F\fi }

\makeatletter
\ifx\CUP@mtlplain@loaded\undefined
\else
\fi
\makeatother
 \makeatletter
 \ifx\CUP@mtlplain@loaded\undefined
   \font\tenbmi=cmmib10 at 10pt
   \font\sevenbmi=cmmib10 at 7pt
   \font\fivebmi=cmmib10 at 5pt

   \newfam\bmifam
   \textfont\bmifam=\tenbmi
   \scriptfont\bmifam=\sevenbmi
   \scriptscriptfont\bmifam=\fivebmi
   
 \fi
 \makeatother
\ifnfsstwo

\fi
\ifnfssone

\fi
\ifoldfss

\fi

\mathchardef\varLambda="0103

\makeatletter
\ifx\CUP@mtlplain@loaded\undefined
\else
\fi
\makeatother
\makeatletter
\ifx\CUP@mtlplain@loaded\undefined
  \font\tenbms=cmbsy10
  \font\sevenbms=cmbsy10 at 7pt
  \font\fivebms=cmbsy10 at 5pt
  \newfam\bmsfam
  \textfont\bmsfam=\tenbms
  \scriptfont\bmsfam=\sevenbms
  \scriptscriptfont\bmsfam=\fivebms

  \edef\bsy@{\hexnumber\bmsfam}
  \mathchardef\bnabla="0\bsy@72
\fi
\makeatother
\def\eg{{e.g.\ }}

\def\etal{\mbox{\em et~al.}}
\def\arcmin{\hbox{$^{\prime}$}}
\def\arcsec{\hbox{$^{\prime\prime}$}}

\title[Gravitational Lensing]{Gravitational Lensing and the
Extragalactic Distance Scale}

\author[R. D. Blandford \& T. Kundi\'c]
{R\ls O\ls G\ls E\ls R\ns D.\ns B\ls L\ls A\ls N\ls D\ls F\ls O\ls
R\ls D\ns
\and \ns
T\ls O\ls M\ls I\ls S\ls L\ls A\ls V\ns K\ls U\ls N\ls D\ls I\ls {\'C}}  
\affiliation{Theoretical Astrophysics, California Institute of
Technology, MC 130-33, Pasadena, CA 91125}

\begin{document}
\ifnfssone
\else
  \ifnfsstwo
  \else
    \ifoldfss
      \let\mathcal\cal
      \let\mathrm\rm
      \let\mathsf\sf
    \fi
  \fi
\fi

\maketitle

\begin{abstract}

The potential of gravitational lenses for providing direct, physical
measurements of the Hubble constant, free from systematic errors
associated with the traditional distance ladder, has long been
recognized. However, it is only recently that there has been a
convincing measurement of a time delay sufficiently accurate to carry
out this program.  By itself, an accurate time delay measurement does
not produce an equivalently definite Hubble constant and the errors
associated with models of the primary lens, propagation through the
potential fluctuations produced by the large-scale structure and the
global geometry of the universe must also be taken into account.  The
prospects for measuring several more time delays and the feasibility
of making the corresponding estimates of the Hubble constant with
total error smaller than ten percent are critically assessed.

\end{abstract}

\firstsection % if your document starts with a section,

\section{Introduction}

Ever since the prescient work of Refsdal (1964, 1966), extragalactic
astronomers have known that a determination of the time delay in the
variation of a multiply imaged quasar could produce a measurement of
the size and age of the universe.  The first definite example of
multiple imaging was discovered in the ``double quasar'' 0957+561A,B
(Walsh, Carswell \& Weymann 1979), and, as it seemed that this dream
might be realized, several groups began monitoring it in the hope of
carrying out this program.\footnote{In fact, they were doing this
under false pretenses because the original estimate of the time delay
was five times too large, due to an uncharacteristic slip in Young
\etal\ (1980).} It took 17 years for a universally accepted time delay
to be measured in this system (Kundi\'c 1996), after a long and
controversial series of papers on the subject (for a review, see
Haarsma \etal\ 1996). The effort in constraining the models of the
lensing mass distribution has paralleled the time delay observations,
resulting in a robust model of the system (Grogin \& Narayan
1996). Many other promising lenses have been discovered since, and
there is now considerable optimism that Refsdal's technique, like many
others discussed at this meeting, is ripe for exploitation. In this
brief review, we will try to summarize recent developments in the
study of gravitational lensing insofar as they are relevant to
measurement of the Hubble constant.  For more detail on the history
and the basic theory, the reader is referred to Blandford \& Narayan
(1992); Schneider, Ehlers \& Falco (1992); and Narayan \& Bartelmann
(1996). Recent observational and theoretical developments are
presented in the 173rd IAU Symposium proceedings (Kochanek \& Hewitt
1996).

\section{The method}

Gravitational lensing relies upon the the propensity of light to
follow null geodesics in curved spacetime.  This means that if we try
to fit the round peg that is the curved space around a mass $M$ into
the square hole that is the pre-relativity framework of Euclid,
Newton, and Kant, then light rays will appear to be deflected through an
angle
\begin{equation}
\hat{\alpha}=\frac{4GM}{bc^2}
\label{alphan.eq}
\end{equation}
where $b$ is the impact parameter. Now, an elementary calculation of
this deflection angle, analogous to the calculation of the deflection
of an ultrarelativistic electron passing by an atomic nucleus, gives
precisely half this deflection.  When one analyzes the general
relativistic calculation, one finds that the other half of the
deflection is directly attributable to the space curvature and that
this is a peculiar prediction of general relativity.  This prediction
has been verified with a relative accuracy of $\sim 0.001$ in a
measurement of the solar deflection (Lebach \etal\ 1985) and we can
take this part of the theory for granted.

It turns out to be quite useful (and rigorously defensible), to use
the Newtonian framework to think in terms of this ``deflection'' and
to treat space as flat but endowed with an artificial
refractive index
\begin{equation}
n=1-\frac{2\Phi}{c^2}.
\label{refind.eq}
\end{equation}
where $\Phi(\vec{\boldmath r})$ is the conventional Newtonian
gravitational potential which $\rightarrow 0$ as $r \rightarrow
\infty$ (Eddington 1919). (As $\Phi<0$, the refractive index exceeds
unity and rays are deflected toward potential wells.)  This also
implies that when photons travel along a ray, they will appear to
travel slower than {\em in vacuo} and will take an extra time $\Delta
t_{\rm grav}$ to pass by a massive object, where
\begin{equation}
\Delta t_{\rm grav}=-\frac{2}{c^3}\int ds \; \Phi
\label{tgrav.eq}
\end{equation}
and where the integral is performed along the ray. (This effect is
also known as the ``Shapiro delay'' and it has been measured with a
fractional accuracy $\sim0.002$ in the solar system by Reasenberg
\etal\ 1979). If the deflector is at a cosmological distance (at a
redshift $z_{\rm d}$), the gravitational delay measured by the
observer will be $(1 + z_{\rm d})$ times longer, where the expansion
factor takes into account the lengthening of time interval
proportional to the lengthening of wave periods.

\begin{figure}
\epsfxsize=\hsize \epsfbox{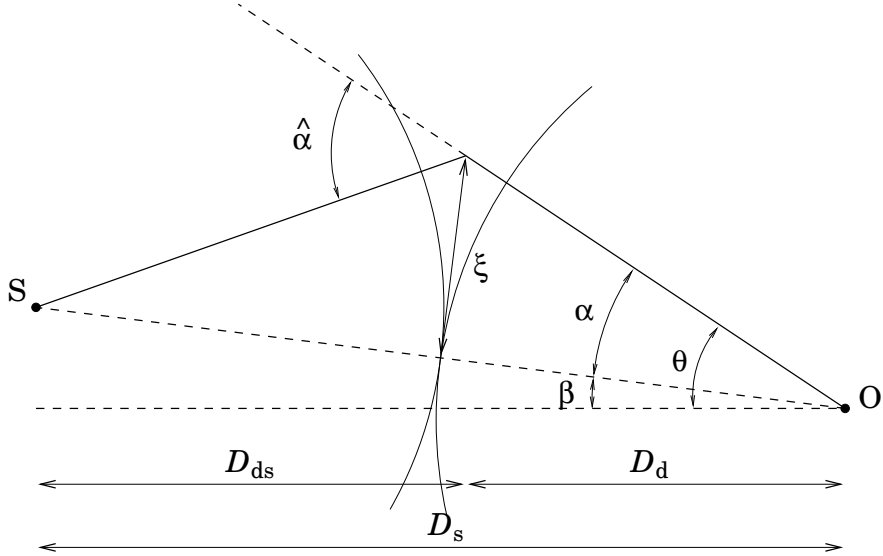}
\caption{
  Illustration of the geometrical time delay in a simplified
  lensing geometry. The two wavefronts shown in the figure
  represent one wave emanating from the source and the other wave
  emanating from the observer backward in time. They meet tangentially
  at the deflector along the {\em undeflected} ray.  The separation
  between these two wavefronts along the {\em deflected} ray is given
  by the expression $\vec\xi \cdot \vec{\hat\alpha}/2$, regardless of
  the space curvature.
  \label{lensgeom.fig}
}
\end{figure}

There is a second, geometrical contribution to the total time delay.
In order to compute this contribution, it is necessary to take account
of the fact that the global geometry of the universe is not
necessarily flat.  Fortunately, this can be done by defining an {\em
angular diameter distance} (\eg Weinberg 1972), which is the ratio of
the proper size of a small source at the time of emission to the
angle that it subtends at a distant observer.  (In computing the
angular diameter distance, it is necessary to allow for the fact that
the universe expands as the light propagates from the source to the
observer.) Now define angular diameter distances from the observer to
the deflector and the source by $D_{\rm d}$, $D_{\rm s}$ respectively
and from the deflector to the source by $D_{\rm ds}$. If we compare
the true deflected ray with the unperturbed ray in the absence of the
deflector, then elementary geometry tells us that the separation of
the two rays at the deflector is given by $\vec\xi=D_{\rm d} D_{\rm
ds} \vec{\hat\alpha}/D_{\rm s}$ (Fig.~\ref{lensgeom.fig}).  Now
imagine two waves, one emanating from the source at the time of
emission, the other emanating from the observer backward in time
leaving now and let these two waves meet tangentially at the deflector
along the undeflected ray. The extra geometrical path is simply the
separation of these wavefronts at the deflector along the deflected
ray, a distance $\vec\xi$ from the undeflected ray. As the rays are
normal to the wavefronts at the deflector, we see that the geometrical
path difference at the deflector is just
$\vec\xi\cdot\vec{\hat\alpha}/2$.  Again, we must multiply by
$(1+z_{\rm d})$.  The net result is an expression for the geometrical
time delay
\begin{equation}
\Delta t_{\rm geom}=(1+z_{\rm d})\frac{D_{\rm d} D_{\rm ds}}{2D_{\rm
s} c} \; \hat\alpha^2 \quad .  
\label{tgeom.eq}
\end{equation}
The total time delay is the sum of the gravitational and the
geometrical contributions.

Both the gravitational and the geometrical time delays are of
comparable magnitude $\sim \hat\alpha^2/H_0$ for cosmologically
distant sources.  For typical deflections $\sim 1 \arcsec$, this would
lead to an estimate $\Delta t \sim 1$~yr.  Curiously, for many of the
sources in which we are most interested, it turns out that the actual
delays are much smaller than this estimate (by up to three orders of
magnitude).  This is because the gravitational and geometrical time
delays tend to cancel each other out and because we tend to
select observationally highly magnified examples of gravitational
lensing in which the image arrangement is quite symmetrical.

In order to evaluate Eqs.~\ref{tgrav.eq}~and~\ref{tgeom.eq} in a given
system, we must construct a model of the deflector.  For a general
mass distribution, the deflection angle, (which can be regarded as a
two dimensional vector as long as it is small), is given by
\begin{equation}
\vec{\hat{\alpha}}=\frac2{c^2}\int ds \; \vec n\times(\vec n \times
\bnabla \Phi) 
\label{deflec}
\end{equation} 
where $\vec n$ is a unit vector along the ray. We can now use this
deflection to solve for the rays. Let us measure the angular position
of the undeflected ray as seen by the observer, relative to an
arbitrary origin on the sky, by $\vec\beta$ and the position of the
deflected ray by $\vec\theta$ (Fig.~\ref{lensgeom.fig}).  These angles
must satisfy the general lens equation
\begin{equation}
\vec\beta=\vec\theta-\vec\alpha(\vec\theta)
\label{lens.eq}
\end{equation} 
where 
\begin{equation}
\vec\alpha=\frac{D_{\rm ds}}{D_{\rm s}} \; \vec{\hat\alpha}
\label{hatalpha.eq}
\end{equation}
$\vec\alpha$ is the {\em reduced} deflection angle.
When Eq.~\ref{lens.eq} has more than one solution, we have a {\em strong}
gravitational lens producing multiple images.  Common strong lenses 
form two or four images.
We can use use Eq.~\ref{lens.eq} to write the total time delay in the
form 
\begin{equation}
\Delta t=K\left[\frac{(\vec\theta-\vec\beta)^2}2-\Psi(\vec\theta)\right]
\label{tgg.eq}
\end{equation}
where 
\begin{equation}
K=(1+z_{\rm d}) \; \frac{D_{\rm d}D_{\rm s}}{D_{\rm ds}c}
\label{kdef.eq}
\end{equation}
and 
\begin{equation}
\Psi=\frac{2D_{\rm ds}}{D_{\rm d}D_{\rm s}c^2} \int ds \; \Phi
\label{psidef.eq}
\end{equation}
is the
{\em scaled surface potential} which satisfies the two-dimensional Poisson
equation 
\begin{equation}
\nabla^2\Psi=\frac{2\Sigma(\vec\theta)}{\Sigma_c} \quad ,
\label{poisson.eq}
\end{equation}
and where
\begin{equation}
\bnabla\Psi=\vec\alpha
\label{alphadef.eq}
\end{equation}
is the reduced deflection angle.  $\Sigma$ is the surface density in
the lens plane and $\Sigma_c=c^2D_{\rm s}/4\pi GD_{\rm ds}D_{\rm d}$
is the so-called {\em critical density}. The derivatives are
performed with respect to $\vec\theta$.

Next, we must calculate the observed magnification. Suppose that we
have a small but finite source that is resolved by the observer. In
the absence of the deflector, the source will appear at position
$\vec\beta$ on the sky.  After deflection, the $\beta$ plane will be
mapped onto the $\theta$ plane.  The Hessian tensor
\begin{equation}
H_{ij}=\frac{\partial\beta_i}{\partial\theta_j}=\delta_{ij}-
\Phi_{,ij}
\label{hessian.eq}
\end{equation}
relates the source to the image.  (Note that, as the deflection is
itself the gradient of a potential, this tensor is symmetric and only
has three independent components.) It is usual to relate the image to
the source and this requires the {\em magnification tensor} which is
the inverse of Eq.~\ref{hessian.eq}:
\begin{equation}
\mu_{ij}=\frac{\partial\theta_i}{\partial\beta_j}=H_{ij}^{-1} \quad .
\end{equation}
This magnification tensor can be decomposed into an isotropic {\em
expansion} and a trace-free pure {\em shear}.  As it is symmetric,
there is no rotation.

Now the usual scalar magnification, denoted by $\mu$, is the ratio of
the flux observed from an unresolved source seen through the deflector
to the flux that would have been measured in the absence of the
deflector. As the intensity is unchanged by the deflector, this ratio
is simply the ratio of the solid angles subtended by the ratio of the
flux with and without lensing, given by the Jacobian
\begin{equation}
\mu=\left|\frac{\partial\theta_j}{\partial\beta_i}\right|=|\mu_{ij}|
\quad .
\end{equation}

These magnifications are not directly observable. Rather, it is the
ratio of the magnifications of separate images of the same source that
one measures.  (Of course this may have to be done at different times
of observation if the source varies so as to make the comparison at
the same time of emission.)  Similarly, if we are able to resolve
angular structure in multiple images of a compact source, for example
using VLBI, then we can also measure the relative magnification tensor
relating two images, $A, B$
\begin{equation}
\mu^{AB}_{ij}=\mu^A_{ik}(\mu^B_{kj})^{-1}.
\label{muAB.eq}
\end{equation}
This tensor need not be symmetric.

The procedure for estimating the Hubble constant then consists of
using the observed positions and magnifications of multiple images of
the same source to construct a model of the imaging geometry which
allows us to deduce $\vec\beta$ for the sources and $\Psi$ for all the
images.  The total time delay for each image can then be computed in
the model up to a multiplicative, redshift-dependent factor $K$ given
by Eq.~\ref{kdef.eq}, which is inversely proportional to $H_0$.  If
the time lags between the variation of two (or more) images can be
measured, it is then possible to get an estimate of $H_0$.

\section{Measuring time delays}

Consider a variable source which produces a variation that can be
observed using more than one ray.  As the travel time along these rays
differs, the corresponding images will vary in brightness at different
observing times.  If the source is continuously variable, then we
should be able to measure the time delay using cross-correlation
techniques. The precision with which it can be measured depends on the
amplitude and timescale of the source variability, on the frequency of
observations and photometric accuracy, and on the value of the time
delay itself.

Let us consider optical observations of quasars. Assuming that quasar
variability is a stationary process, it can be described in terms of
the first-order structure function:
\begin{equation}
V(\tau) = \langle [ m(t + \tau) - m(t) ]^2 \rangle \quad, 
\end{equation}
where $m(t)$, $m(t + \tau)$ are quasar magnitudes recorded at two
epochs separated by an interval $\tau$ (Simonetti, Cordes \& Heeschen
1985). The qualifier ``stationary'' refers to the assumption that $V$
is not a function of the observation epoch $t$, but only of the time
lag $\tau$ between two points on the light curve.  In practice,
$V(\tau)$ can be well approximated by a power law for time delays
between a few days and a few years:
\begin{equation}
[V(\tau)]^{1/2} \sim 0.015 \; {\rm mag} \; \left( \tau/{\rm day}
\right)^{0.35} \quad,
\label{structure.eq}
\end{equation}
where the numerical coefficients were estimated from Press, Rybicki \&
Hewitt (1992a); Cristiani \etal\ (1996); and from the light curves of
several lenses monitored at the Apache Point Observatory. The
amplitude of $V(\tau)$ depends on the bandpass and it is larger at
bluer wavelengths.

Equation~\ref{structure.eq} illustrates the difficulties associated
with measuring short time delays. If the predicted delay is a few
days, the expected amplitude of quasar variations is only $\sim 0.02$
magnitudes, requiring millimagnitude photometry in systems that are
often complicated and marginally resolved from the ground (In
quadruple lenses, one has to resolve four quasar images in a $\sim
1\arcsec$ radius, superimposed on the diffuse light of the lensing
galaxy). Even in a resolved system with a much longer time delay, the
double quasar 0957+561, there has been much debate about the correct
value of the delay, with estimates ranging from $\sim 540$ days
(Leh\'ar \etal\ 1992; Press \etal\ 1992a, 1992b; Beskin \& Oknyanskij
1995) to $\sim 420$ days (Vanderriest \etal\ 1989; Schild \& Cholfin
1986; Schild \& Thomson 1995; Pelt \etal\ 1994, 1996). Optical data
recently acquired at the Apache Point Observatory by Kundi\'c \etal\
(1995, 1996) unambiguously measure a delay of $\Delta t = 417 \pm
3$~days, vindicating the second group. This result required $\sim 100$
flux measurements in two observing seasons with a median photometric
error of $< 1$\%.

A less accurate time delay of $12\pm3$~days has been reported for the
radio ring B0218+357 on the basis of radio polarization measurements
(Corbett \etal\ 1996). The advantage of the polarization method over
direct photometry is that only one parameter (time delay) is needed to
align light curves of two images. Alignment of photometric light
curves requires an additional parameter, corresponding to the relative
magnification of the two images. More recently, Schechter \etal\
(1996) reported a measurement of two time delays in the quadruple
quasar PG1115+080: $23.7\pm3.4$~days between images B and C and
9.4~days between images C and A. Unfortunately, it is not yet possible
to model PG1115+080 accurately and so this impressive observation
cannot furnish a useful estimate of the Hubble constant at this
time. Sources like B1422+231 (Hjorth \etal\ 1996) and PKS1830-211 (van
Ommen \etal\ 1995) are also currently being monitored.

\section{Modeling strong galaxy lenses}

\subsection{The models}

Any estimate of $H_0$ derived from gravitational lensing will be no
better than the lens model that is deduced from observations of the
positions and fluxes of the individual images.  Lens modeling is,
necessarily, a rather subjective business.  The procedure that has to
be followed depends upon whether or not the images are resolved.  In
most cases, there are $N$ unresolved point images giving $2(N - 1)$
relative coordinates.  There are also $(N - 1)$ relative fluxes which
can be fit to the magnification ratios.  Since sources are variable,
fluxes must be compared at the same emission time.  (A complication
that can arise at optical wavelengths is that individual stars in the
lensing galaxy can cause additional, variable magnification, called
{\em microlensing}, if the source is sufficiently compact.  However,
this is unlikely to be a problem at infrared or radio wavelengths.)
So, with $N$ point images, there are usually $3(N - 1)$ observables
which can be used in model fitting. Once the relative time delays in
the system are measured, they provide $(N - 2)$ additional
constraints.

Resolved images contain more information.  Firstly, when compact radio
sources are resolved using VLBI, the individual images ought to be
related by a simple, four parameter, linear transformation
(Eq.~\ref{muAB.eq}).  This can be measured and is more useful than the
flux ratio alone in constraining the model, and would give an
additional $3 (N - 1)$ observables.  Secondly, if the radio structure
is large enough, it is possible to expand to one higher order and
measure the gradient of $\mu_{ij}$.  Thirdly, some sources are so
extended that they form ring-like images.  This can happen at either
radio or optical wavelengths.  In this case we have a large number,
perhaps thousands of independent pixels to match up.  In principle,
this can lead to a highly constrained lens (and source) model.  The
best techniques for tackling this problem at radio wavelengths
(Wallington, Kochanek \& Narayan 1996) incorporate the model fitting
as a stage in the map making procedure and, in some cases this can
lead to impressively good models.

In order to model a simple gravitational lens, the scaled potential
$\Psi$ of the lensing galaxy, (plus any additional galaxies close
enough to contribute to the deflection) is modeled using a simple
function that contains adjustable parameters that measure the depth of
the potential, its core radius, its ellipticity and its radial
variation.  A convenient functional form is the {\em elliptical potential}
\begin{equation}
\Psi_e(\vec\theta)=b[s^2+(1-\gamma_c)x^2-2\gamma_s xy+(1+\gamma_c)y^2]^q
\label{elliptical.eq}
\end{equation}
where $\vec{\theta}=(x,y)$ measures distance from the center of the
potential which is located at $(x_0,y_0)$.  The two parameters
$\gamma_c,\gamma_s$ combine to describe the ellipticity and the
position angle of the major axis.  The exponent $q$ measures the
radial variation of the density at large radius.  A value $q=0.5$
corresponds to an isothermal sphere and a value $q=1$ has
asymptotically $\Sigma\propto r^{-2}$ as might be appropriate if, for
example, mass traces light as in a Hubble profile.  The elliptical
potential has the advantage that it is quick to compute and good for
searches in multi-dimensional parameter space.  Its drawback is that
the corresponding surface density is peanut-shaped at large
ellipticity. Either two potentials must be superposed to generate
realistic surface density distributions or a choice made from a
smorgasbord of more complicated potentials (\eg Schneider \etal\
1992).  In some cases, these potentials can be located and oriented
using the observed galaxy image. However, we know that the mass in
galaxies does not completely trace the light; in particular it
diminishes more slowly with increasing radius.  Therefore, we still
have some freedom in modeling the mass distribution of lensing
galaxies, even if their light profiles are well known.  An example of
the elliptical potential model is shown in
Fig.~\ref{1608.fig}. Motivated by the optical HST images of 1608+656,
two elliptical potentials are superimposed to produce the observed
quad configuration.

It is also conventional to augment the lensing galaxy potential with a
simple quadratic form,
\begin{equation}
\Psi_q(\vec\theta)=-\Gamma_c(x^2-y^2) -2\Gamma_s xy \quad ,
\label{quadratic.eq}
\end{equation}
which is trace-free and corresponds to a pure shear of the images.
The external shear has two origins: dark matter lying in the lens
plane associated with galaxies outside the strong lensing region and
large-scale structure along the line of sight, as we discuss in more
detail below.

A common procedure for modeling well observed gravitational lenses is
to decide upon a list of independent observables and a smaller number
of lens model parameters. These parameters are then varied so as to
minimize some measure of the goodness of fit, typically a $\chi^2$
associated with the differences in the observables as predicted by the
model from those actually measured.  This approach has the merit of
rewarding simplicity.  However, the fit is usually somewhat imperfect
and it is difficult to assign a formal error.  A somewhat better
approach, that is much harder to implement in practice, is to
construct families of more elaborate models that may contain more
parameters than observables and which are consequently
underdetermined.  Using this approach, it ought to be possible to
derive many models that exactly recover the observables.  We can then
assign two types of error to the model time delay. One is associated
with the measurement errors in the observables, the other depends upon
the freedom in the models and it is here that the outcome depends most
subjectively upon what we are prepared to countenance.

\subsection{Modeling ``quad'' sources}

Let us describe some of the dangers and possibilities involved in
model fitting by considering image formation in a simple case, the
so-called ``quad'' sources.  These are quadruply imaged point sources
formed by a galaxy for which the circular symmetry is broken by an
elliptical perturbation.  (Depending upon the nature of the potential
in the galaxy nucleus, there may also be a fifth, faint, central image
which we shall ignore.)  For illustration, we use the simplest
possible model of a singular isothermal sphere potential perturbed by
a quadratic shear, and work to first order in the ellipticity ({\em
cf} Blandford \& Kovner 1988).

\begin{figure}
\epsfxsize=\hsize \epsfbox{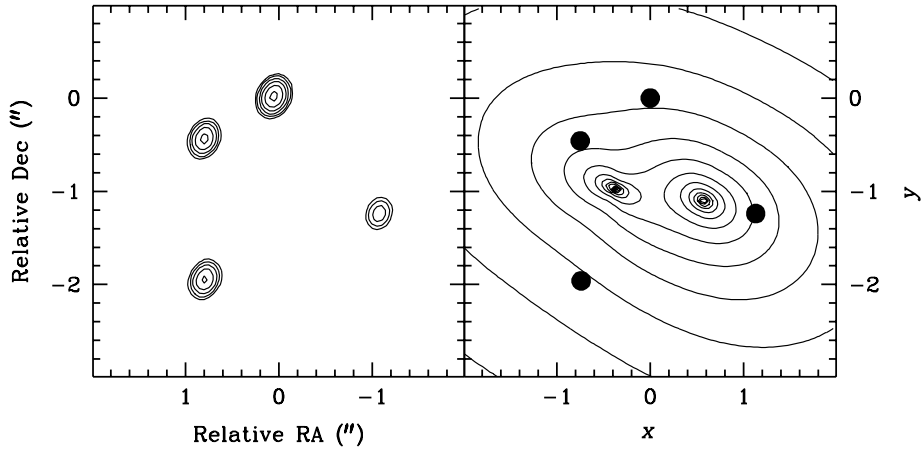}
\caption{
  {\bf Left:} Radio map of the gravitational lens system
  1608+656. The system was observed on 1996 October 10 with the VLA
  at 8.4 GHz. The map shows four unresolved images of the background
  source (at $z_{\rm s}=1.39$). The brightest image (A) is located at the
  origin. (Courtesy of Chris Fassnacht.)
  {\bf Right:} A simple model for 1608+656, consisting of two
  elliptical potentials. Solid lines represent logarithmically spaced
  density contours of the lensing mass distribution. The observed 
  image positions (marked with dots) and their relative fluxes are
  accurately reproduced by this model. 
  \label{1608.fig}
}
\end{figure}

\begin{figure}
\epsfxsize=\hsize \epsfbox{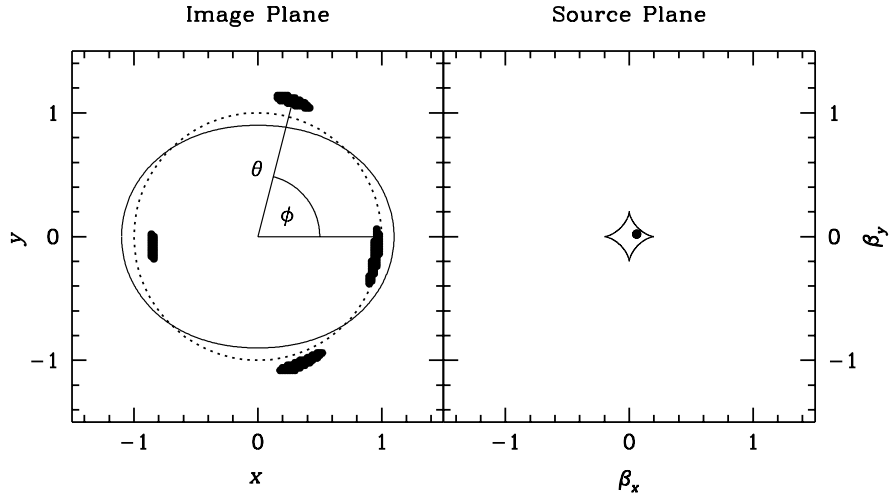}
\caption{
  Lensing geometry giving rise to a four-image (``quad'')
  configuration. If the source is located within the astroid (diamond)
  caustic produced by an elliptical potential (right panel), there
  will be four tangentially elongated images near the critical
  curve (solid line) in the image plane (left panel). Dotted line in
  the left panel corresponds to the Einstein ring of the unperturbed,
  circularly symmetric lens. 
  \label{quad.fig}
}
\end{figure}

We write the potential in the form
\begin{equation}
\Psi_s (\vec\theta) = \theta_0 \theta - \frac{1}{2} \epsilon \theta^2
\cos 2\phi \quad , 
\label{quadpot.eq}
\end{equation}
where polar coordinates $\vec\theta = (\theta,\phi)$ refer to the
center of the potential located at the center of the observed lensing
galaxy (Fig.~\ref{quad.fig}). The first term $\Psi_s^0=\theta_0\theta$
describes an isothermal sphere ($\Sigma \propto \theta^{-1}$), and is
circularly symmetric. The second term is the non-circular
perturbation. We assume that we know its orientation from the shape of
the galaxy but treat the ellipticity as unknown.  If we set $\epsilon
= 0$ and place a source on the axis, then the image will be a circular
ring of radius $\theta_0$ known as the {\em Einstein ring}. We now
switch on the perturbation and solve for the source position to linear
order assuming that $\epsilon \ll 1$. We find that the source position
is
\begin{equation}
\vec\beta = \left [ \delta + \epsilon \theta_0 \cos 2\phi \right]
\vec{s} - \epsilon \theta_0 \sin 2\phi \; \vec{t} \quad , 
\label{quadbeta.eq}
\end{equation}
where $\delta = \theta - \theta_0$, and $\vec s, \, \vec t$ are unit
vectors in the radial and tangential directions respectively.  We can
now evaluate the Hessian and find that its eigenvalues are
\begin{equation}
h_1 = 1 \qquad h_2 = \frac{\delta}{\theta_0} - \epsilon \cos 2\phi
\quad , 
\label{quadhess.eq}
\end{equation}
and the principal axes are rotated with respect to $\vec s, \, \vec t$
by an angle 
\begin{equation}
\psi = -\epsilon \sin 2\phi \quad .
\label{quadpsi.eq}
\end{equation} 
These eigenvalues are the reciprocals of the quasi-radial and
quasi-tangential magnifications, so that the total magnification is
$\mu = (h_1 h_2)^{-1} = ({\delta}/{\theta_0} - \epsilon \cos
2\phi)^{-1}$. The locus of the infinite magnification on the sky, the
``critical curve'' is then given by $h_2 = 0$ or
\begin{equation}
\delta_c = \epsilon \theta_0 \cos 2\phi \quad .
\label{quaddelt.eq}
\end{equation}
The critical curve is the image of the ``caustic'', the locus in the
source plane of points that are infinitely magnified.  Its equation is
\begin{equation}
\vec\beta_c = [\beta_{cx}, \, \beta_{cy}]
            = 2 \epsilon \theta_0 [\cos^3\phi, \, -\sin^3\phi] \quad ,
\label{quadbetc.eq}
\end{equation}
which is the parametric equation of an astroid.  If we know the source
position, $\vec\beta$, and it is located within the astroid then there
will be four images located close to the critical curve
(Fig.~\ref{quad.fig}). If the source lies outside the astroid, there
will be two images. 

This model has two free parameters, $\epsilon$ and $\theta_0$, and as
we may have eleven observables in a four-image lens (three pairs of
relative positions + three flux ratios + two time delay ratios), there
are many internal consistency checks. For example, the angles of the
four images must satisfy
\begin{equation}
\left [ \sum_{i = 1}^{4} \cos \phi_i \right ]^2 = -4 \prod_{i = 1}^{4}
\cos \phi_i \quad .
\label{quadcons.eq}
\end{equation}
Now, in order to trust the model, we contend that it is necessary that
every observable be fit within its measurement error, or there should
be a good reason why it can be excused. For example, we might have a
spectroscopic indication that microlensing is at work in one of the
images and then we would discount the measurement of its flux.

Now, it is extremely unlikely that any lens will ever be found that is
this simple. So, after we are sure that there is inconsistency, we
must introduce additional parameters or change the model. Following
the above approach of expanding about the Einstein ring, we can (or
may have to) allow the center of the potential to vary, adjust the
orientation of the perturbations, allow the ellipticity $\epsilon$ to
vary with radius $\theta$, introduce higher harmonics (\eg $\propto\cos
4\phi$ for box-like equipotentials) and so on. Alternatively, we can
turn to a density-based model. The nature and quality of the observations
will usually dictate this choice.

What we discover when we carry out this procedure is that the allowed
range of observables is quite sensitive not just to the parameters,
but to the functional form of the model. For example,
constraint~\ref{quadcons.eq} is quite general but can be violated if
(not unreasonably) we were to add a $\cos 4\phi$ term. In addition,
the shape of the unperturbed, circular potential $\Psi^0_s$ controls
the quasi-radial magnification and changing it changes $h_1$ to $1 -
d^2\Psi^0_s/d\theta^2$. The ratios of time lags are fairly well fixed
by the observed image magnifications and location; however, the
absolute values (in which we are primarily interested) are extremely
sensitive to radial variation of the ellipticity and this can be
measured if we have very accurate image positions or an extended
source. Finally, we should probably always include the quadratic form
Eq.~\ref{quadratic.eq} to take account of distant mass.

There are further possible consistency checks.  For example, if we can
make VLBI observations of the individual image components, as should
be possible for 1608+656 (Myers \etal\ 1995, Fassnacht \etal\ 1996;
Fig.~\ref{1608.fig}), and resolve structure in the radial direction, we
can also check both the radial eigenvalue $h_1$, and the orientation
angle $\psi$ for each of the images.  Even better, if we have more
than one source, as appears to be the case in 1933+503 (Sykes \etal\
1996), than we have two independent opportunities to test the model
and derive the parameters.  (This approach may also be of interest for
cluster arcs.)

Let us use the quad H1413+117 for illustration and try to model it
using a simple potential (Eq.~\ref{quadpot.eq}).  We can guess the
lens center pretty accurately, but not the position angle of the major
axis of the elliptical perturbation, $\phi_0$. Using
Eq.~\ref{quadcons.eq}, we find four choices for $\phi_0$ and we
examine each of these in turn to try to find a consistent solution for
the source position $\vec\beta$ by varying $\epsilon$, $\theta_0$, and
the lens center. We find that none of our choices of $\phi_0$ can
accomplish this, although one comes close. However, if we now compute
the magnification ratios, we find that they are quite inconsistent
with the observations. Therefore we can assuredly reject our lens
model. It turns out that it is possible to reproduce the positions and
fluxes using an elliptical potential with parameters $\{b, \, s, \, q,
\, \gamma_c, \, \gamma_s, \, x_c, \, y_c\}$ offset by an external
shear with parameters $\Gamma_c,\,\Gamma_s$. Although this seems the
most plausible model, it is not unique (Yun \etal\ 1996, preprint) and
this would be a concern were this a good choice for
monitoring. Unfortunately, it is not as the predicted time delays are
almost certainly too small.

\subsection{Uncertainties in the models}

An important question that we must now address is ``How do we assign a
formal error to the predicted model on the basis of a deflector
model?'' One procedure is to define a $\chi^2$ using careful values
for the measurement errors and then choose parameters $p_i$ that
minimize this statistic.  We then compute a covariance matrix
\begin{equation}
V_{ij}={\partial^2\chi^2\over\partial p_i\partial p_j}
\label{covmatdef.eq}
\end{equation}
and invert and diagonalize this quantity so as to form the error
ellipsoid.  We next compute the maximum fractional change in the time
delay within this error ellipsoid and quote this as the fractional
error in the derived Hubble constant varying the parameters
independently until $\chi^2/\nu$ increases by unity.  The maximum
change in the derived value of the delay as we go through this
procedure is the required error in the delay.

However, this is still not enough because existence does not imply
uniqueness.  Suppose that we have a model that truly reproduces all
the observables giving an acceptable value for $\chi^2$ and for which
the underlying mass distribution is dynamically possible. We should
still aggressively explore all other classes of models that can also
fit the observations but yet which produce disjoint estimates for the
time delay.  The true uncertainty in the Hubble constant is given by
the union of all of these models.  This is a large task.  Clearly, we
may have to introduce some practical limitations.  It is probably
quite safe to reject dynamically unreasonable mass distributions, for
example positive radial gradients in surface density, but are we
allowed to posit isolated concentrations of unseen mass? Of course,
only time will tell whether there are genuinely dark galaxies, but
until we can demonstrate that all significant perturbers are luminous
we must allow for this possibility.

There is already an indication that there may be more unseen mass than
we have already included in the discovery that the incidence of quad
sources relative to doubles is much greater than one might have
expected if the lenses are about as elliptical as normal galaxies
(King \& Browne 1996).  Three explanations have been advanced for this
discrepancy.  Firstly, the total mass associated with the observed
lensing galaxies may be highly elliptical, or, indeed, quite
irregular.  (It appears that observed faint galaxies have a median
ellipticity of $\sim0.4$ as opposed to $\sim0.15$ for local galaxies.)
Secondly, many of the lenses may actually have unseen companions so
that the combined potential will generally be quite elliptical.
Thirdly, as we discuss below, propagation effects associated dark
matter long the line of sight may actually promote the formation of
quads.

\subsection{The double quasar 0957+561}

We now briefly turn from quads to the important case of 0957+561, the
only gravitational lens where the time delay has been measured with
high accuracy (Kundi\'c \etal\ 1996). Traditionally, 0957+561 was
thought to be a difficult system to model because of the small number
of constraints in its two-image configuration and because of the
complexity of the lensing potential -- the primary lensing galaxy G1
is surrounded by a cluster.  To a large extent, these problems have
been resolved in the extensive theoretical study of Grogin \& Narayan
(1996). The crucial set of constraints in the GN model was provided by
high spatial resolution VLBI mapping of Garrett \etal\ (1994), which
resolved inner jets in both images of the source into five centers of
emission. Mapping of one jet into the other fixed the relative
magnification tensor of images A and B (Eq.~\ref{muAB.eq}), as well as
its gradients along and perpendicular to the jet. This, in turn,
provided a tight constraint on the radial mass profile ($dM/dr$) at
the image locations, which, together with the total enclosed mass $M(<
r)$, controls the conversion factor between the time delay and the
physical distance to the lens. The importance of the $dM/dr$ term was
nicely illustrated by Wambsganss \& Paczy\'nski (1994).

The remaining model degeneracy in the 0957+561 system between the
lensing galaxy G1 and its host cluster cannot be removed by using 
relative image positions and magnifications (Falco, Gorenstein \&
Shapiro 1985). It has to be broken by either directly measuring the
mass of G1 via its velocity dispersion, or by measuring the surface
density in the cluster from its weak shearing effect on the images of
background galaxies (\eg Tyson, Valdes \& Wenk 1990, Kaiser \& Squires
1993). If both of these parameters are independently determined, they
provide an important consistency check on the model.

Falco \etal\ (1996, private communication) recently observed G1 with
the LRIS spectrograph at the Keck. Their high signal-to-noise
spectrum yields a line-of-sight velocity dispersion roughly in the
range $\sigma_{\rm obs} = 275 \pm 30$ km/s (2$\sigma$), consistent
with the only published measurement by Rhee (1991), $\sigma_{\rm obs}
= 303 \pm 50$ km/s.  Using a deep CFHT image of the field, Fischer
\etal\ (1996) mapped the cluster mass distribution from the distortion
of faint background galaxies. Adopting their estimate of the mean
background galaxy redshift ($z_{\rm b} = 1.2$) gives the dimensionless
cluster surface mass density $\kappa = \Sigma/\Sigma_c = 0.18 \pm
0.11$ (2$\sigma$).  Using these two results and their measurement of
the time delay, Kundi\'c \etal\ (1996) find 
\begin{eqnarray}
H_0 & = & 67^{+10}_{-11} \left( \frac{1 - \kappa}{0.82} \right) \;
          {\rm km \, s^{-1} \, Mpc^{-1}} \\
    & = & 64^{+13}_{-14} \left( \frac{\sigma_{\rm obs}}{275 \, {\rm km
          \, s^{-1}}} \right)^2 \; {\rm km \, s^{-1}\, Mpc^{-1}} \quad .
\end{eqnarray}
The uncertainty in $H_0$ is quoted at 2$\sigma$ and allows for finite
aperture effects and anisotropy of stellar orbits in conversion of
$\sigma_{\rm obs}$ to G1 mass (GN). While this estimate of $H_0$ can
certainly be improved by future observations, it is already 10\%
accurate at the 1$\sigma$ level. 
 
\section{Systematic errors}

There are a variety of possible systematic errors which afflict
gravitational lens determinations of the Hubble constant. The first of
these, which we have already mentioned, is the degeneracy between $H_0$
and a uniform density sheet in the lens plane.  For example, suppose
that there is a uniform density circular disk of matter covering all
the images.  This will act like a simple (Gaussian) converging lens
and bring the same rays that would have met at the observer to a
common focus closer to the lens.  In other words, the potential
variation is quadratic and so the contribution to the gravitational
delay is also quadratic, just like the geometrical delay, from which
it is therefore indistinguishable.  It has been argued that, as any
sheet must have positive mass density, we can only set an upper bound
on the Hubble constant (Falco \etal\ 1985).

A second uncertainty is associated with the choice of cosmographic
world model.  Let us suppose that the universe is of homogeneous
Friedmann-Robertson-Walker (FRW) type so that it is parametrized by
the current density parameter $\Omega_0$.  The angular diameter
distances to high redshift sources and lenses depend quite sensitively
upon $\Omega_0$.  However, the combination $D_{\rm d}D_{\rm s}/D_{\rm
ds}$, relevant for the inferred Hubble constant, is relatively
insensitive.  For example, in the 0957+561 system, $K = 39.9 h^{-1}$
days arcsec$^{-2}$ for an Einstein-de Sitter universe with $\Omega_0 =
1$, and $K = 42.9 h^{-1}$ days arcsec$^{-2}$ for an open universe with
$\Omega_0 = 0.1$. If we take an additional step and introduce a
cosmological constant while keeping the universe flat, the change in
$K$ is even smaller. For $0 < \Omega_\Lambda < 1$, $K$ attains a
maximum of $41.7 h^{-1}$ days arcsec$^{-2}$ in the $\Omega_0=0.25,
\Omega_\Lambda=0.75$ model. In a lens system with higher redshifts,
however, the choice of cosmological model is more important. We can
thus use high-redshift lenses to constrain the density parameter
$\Omega_0$ once $H_0$ has been measured. For example, in the $z_s =
3.62$ lensed BAL quasar B1422+231, the difference in $K$ between
$\Omega_0 = 1$ and $\Omega_0 = 0.1$ models amounts to 24\% (assuming
the lens redshift of $z_d = 0.65$, as reported by Hammer \etal\ 1995).

Now let us turn to inhomogeneous cosmological models.  If the universe
has an overall mean density sufficient to allow it to follow FRW
dynamics on the average but with this mass confined to small
concentrated lumps, none of which intersect the line of sight, then
individual ray congruences will be subject to zero convergence as they
cannot pass through any matter. In this case, the angular diameter
distance must be changed to the {\em affine parameter} (Press \& Gunn
1973). If, for example, the mass in an Einstein-De Sitter universe
were concentrated in large, distant lumps, then $K$ would change by
over 10\% for B1422+231. However, this assumption is quite
unrealistic, because the combined tidal influence of these lumps must,
on the average, reproduce the cosmography of a homogeneous universe,
and, in fact, it does. In what follows we assume that our line of
sight is not special in this sense and that individual dark masses are
small enough that we can treat them as smoothly distributed.

The next type of systematic error is starting to be taken more
seriously and has broader implications.  This is the error introduced
in the measurement of $H_0$ due to large-scale structure distributed
along the line of sight from the source to the observer.  A simple and
standard description of large-scale structure in the universe is to
form the power spectrum of relative density fluctuations, $P(k)$, that
have supposedly grown from perturbations similar to those that we
observe in the microwave background fluctuations.  The short
wavelength perturbations [$k > k_m \sim (10~{\rm Mpc})^{-1}$] have
grown to non-linear strength; the long wavelength perturbations should
still be in the linear regime with $P(k) \propto k$.  In the linear
regime the potential fluctuations therefore satisfy $\delta\Phi\propto
k^{-2}\delta\rho\propto k^{-2}[k^3P(k)]^{1/2}\sim$ constant.  In fact,
the potential fluctuation is not just constant with linear scale but
also with time and has a value $\delta\Phi\sim3\times10^{-5}c^2$ as
normalized to the fluctuations measured by COBE.

The perturbations that are most important for perturbing gravitational
lens images are those for which $k^2P(k)$ is maximized, i.e. with $k\sim
k_m \sim (10~{\rm Mpc})^{-1}$.  As typical angular diameter distances
are $D\sim1$~Gpc, these subtend angles $\sim30'$, much larger than the
strong lensing regions, and we expect to see $N\sim k_mD\sim100$
perturbations, of both signs, along the line of sight.  If we consider
a single ray passing through a roughly spherical perturbation, it will
be deflected through an angle $\delta\alpha\sim\delta\Phi/c^2$ and as
there are $N$ such deflections adding stochastically, the total
deflection will $\delta\theta\sim N^{1/2}\delta\alpha\sim1'$.  (It is
amusing that although that there has been so much trouble taken to
point HST to a small fraction of a pixel, the universe has a pointing
error of about the width of WFPC chip!)  In addition, each fluctuation
will introduce a propagation time fluctuation $\delta\Phi /k_m
c^3 \sim (\delta\Phi/c^2)(D/Nc)$.  These fluctuations add
stochastically giving a total time difference relative to a
homogeneous universe of $\delta t \sim (\delta \Phi /c^2) (D / N^{1/2}
c) \sim 10^4$~yr.  However, none of this matters because there is no
way to detect the deflection or time delay of a single ray.

The situation changes when we consider two rays separated by a small
angle $\theta$ (Blandford \& Jaroszy\'nski 1981).  On passing though a
single perturbation, the angular separation $\theta$ will change by an
angle $\sim k_m(D\theta)(\delta\Phi/c^2)$.  Again, summing the effects
of $N$ such fluctuations stochastically gives an estimate of the total
fractional change in $\theta$, a measure of the strength of both the
magnification and the shear fluctuations. We obtain
\begin{equation}
\left(\frac{\delta\theta}{\theta}\right)\sim\delta\mu\sim\epsilon\sim
\frac{\delta\Phi}{c^2}N^{3/2}\sim 0.03 \quad .
\end{equation}
These fluctuations should be coherent over angles $\sim (1/k_m D )
\sim 30\arcmin$ on the sky.  There have been attempts to measure this
signal using the images of faint galaxies (\eg Mould \etal\ 1994).

There is a subtlety when we consider the observable time fluctuations.
If the separation of a pair of rays $\theta$ is much less than the
angular scale of the fluctuation, we can Taylor expand the potential.
We have already dismissed the constant term and might expect the
linear term to be observable.  However, this is not the case as can be
seen by noting that its effect simply deflects both rays through the
same angle, just like a prism, and does not increase the optical path
at all.  We have to expand $\Phi$ to quadratic order across the rays
to obtain a total time difference due to the fluctuations of $\delta t
\sim (\delta\Phi/c^2) (D/N^{1/2}c) (k_m\theta D)^2 \sim
(\delta\Phi/c^2) N^{3/2} (D\theta^2/c)$.  Now suppose that these two
rays are associated with a common source and observer in a
gravitational lens.  The time delay in an otherwise homogeneous
universe would be $t\sim D\theta^2/c$ and so the relative change in
the arrival time will satisfy
\begin{equation}
\left(\frac{\delta t}t\right) \sim \frac{\delta\Phi}{c^2}N^{3/2} \sim
0.03  \quad ,
\end{equation}
as above. Therefore to order of magnitude, we find that a few
percent error in the derived value of the Hubble constant will be
introduced by the effects of large-scale structure and the effect
deserves computing seriously.

This problem has been addressed by Seljak (1994) and Bar-Kana (1996),
following earlier papers by Kovner (1987) and Narayan (1991). They
find three distinct effects:
\begin{enumerate}
\item The images of the source, their positions and their shapes, are
subject to a global shear transformation. This effect will generally
be absorbed in the quadratic contribution to shear from the
large-scale distribution of mass in the deflector plane.
\item The images of the deflecting galaxies, their positions relative
to the images of the source and their shapes will be subject to an
additional shear.  Naturally this shear is only caused by fluctuations
between us and the deflector. This effect means that there is a cosmic
uncertainty in the image positions which must be taken into account
when we make the models.
\item There will be quadratic potential fluctuations acting on the
unperturbed rays that cause stochastic gravitational delays that
create additional direct changes in the measured lags. These will
translate directly into errors in $H_0$. There is an interesting
observational possibility here. We can contemplate performing deep
redshift surveys in the vicinity of the most promising lenses and
rather than treat these effects as random errors, try to remove them
directly.
\end{enumerate}

\section{Is a ten percent measurement of $H_0$ attainable?}

In order for a gravitational lens measurement of $H_0$ to be of
primary importance, it is probably going to be necessary for it to
have an accuracy $\sim10$~percent.  Is this realistically attainable?
Firstly, we can say that it is possible to measure the time delay to
better than seven percent accuracy as demonstrated by the optical
observations of 0957+561. This should be repeatable in other selected
sources. However, accuracies this good have yet to be demonstrated at
radio wavelengths, partly because variations over the expected time
delays have smaller amplitude. It is prudent to take into account such
factors as the expected delay (neither too short nor too long), the
signal-to-noise ratio in individual flux measurements, and a
measured structure function before deciding which few sources to
monitor regularly. Simulations should be used for deciding how
frequently to measure the flux.

Secondly, we need to be able to model the lenses with a total
fractional uncertainty in the predicted time delay also below $\sim
7$~percent. Here, there are two challenges. As we have discussed, the
model itself must be constrained so well that we are confident that
there are no other models of the deflector that recover all the
observables to within the measurement errors and yet predict seriously
different delays. In addition, we must convince ourselves that the
perturbative effects of the large-scale structure along the line of
sight do not influence our result at this level.

There is a plethora of alternative schemes to measure $H_0$, and each
one of them has enthusiastic advocates.  In this competitive
environment, all methods carry a burden of proof and must demonstrate
a reliability and reproducibility if they are to become widely
accepted.  In the particular case of gravitational lenses, this means
that we must derive consistent values of the Hubble constant in
several, probably at least four, cases.  It is also probably necessary
for the models to be specified and analyzed prior to measuring the
time delay.  The considerations outlined above suggest that this goal
is attainable and, as the gravitational lens method is directly
physical and free from the calibration uncertainties that bedevil most
other methods it is well worth the observational effort to carry out
this program.

\begin{acknowledgments}
We acknowledge support under NSF grants AST~92-23370 and
AST~95-29170. Support for this work was also provided by NASA through
grant number AR-06337.15-94A from the Space Telescope Science
Institute, which is operated by the Association of Universities for
Research in Astronomy, Inc., under NASA contract NAS5-26555. We thank
Ed Turner, David Hogg and the whole CLASS collaboration, especially
Ian Browne, Chris Fassnacht, Sunita Nair and Tony Readhead, for
discussions.
\end{acknowledgments}

\end{document}